\documentclass[%
 reprint,
superscriptaddress,
 amsmath,amssymb,
 aps,pre
]{revtex4-2}

\usepackage{graphicx}
\usepackage{dcolumn}
\usepackage{bm}
\usepackage[breaklinks]{hyperref}
\hypersetup{
    colorlinks=true,
    linkcolor=blue,
    filecolor=blue,
    urlcolor=cyan,
    }
\usepackage{breakurl}
\usepackage[normalem]{ulem}

\usepackage{placeins}
\usepackage{verbatim}

\begin{document}


\title{Network percolation provides early warnings of abrupt changes in
coupled oscillatory systems: An explanatory analysis}

\author{No\'{e}mie Ehstand}
\email{n.ehstand@ifisc.uib-csic.es}
\affiliation{%
IFISC  (CSIC-UIB), Instituto  de  F\'{\i}sica  Interdisciplinar y Sistemas Complejos, Campus Universitat de les Illes Balears, E-07122 Palma de Mallorca, Spain
}%
\author{Reik V. Donner}%
\affiliation{
Department of Water, Environment, Construction and Safety, Magdeburg-Stendal University of Applied Sciences, Breitscheidstra{\ss}e 2, D-39114 Magdeburg, Germany
}
\affiliation{ Research Department IV-Complexity Science and Research Department I-Earth System Analysis, Potsdam Institute for Climate Impact Research (PIK) -- Member of the
Leibniz Association, Telegrafenberg A31, D-14473 Potsdam, Germany
}%

\author{Crist\'{o}bal L\'{o}pez}
\affiliation{%
IFISC  (CSIC-UIB), Instituto  de  F\'{\i}sica  Interdisciplinar y Sistemas Complejos,
Campus Universitat de les Illes Balears, E-07122 Palma de Mallorca, Spain
}%

\author{Emilio Hern\'{a}ndez-Garc\'{\i}a}
\affiliation{%
IFISC  (CSIC-UIB), Instituto  de  F\'{\i}sica  Interdisciplinar y Sistemas Complejos,
Campus Universitat de les Illes Balears, E-07122 Palma de Mallorca, Spain
}%

\date{\today}

\begin{abstract}
Functional networks are powerful tools to study statistical
interdependency structures in spatially extended or
multivariable systems. They have been used to get insights
into the dynamics of complex systems in various areas of
science. In particular, percolation properties of correlation
networks have been employed to identify early warning signals
of critical transitions. In this work, we further investigate
the corresponding potential of percolation measures for the
anticipation of different types of sudden shifts in the state
of coupled irregularly oscillating systems. As a paradigmatic
model system, we study the dynamics of a ring of diffusively
coupled noisy FitzHugh-Nagumo oscillators and show that, when
the oscillators are nearly completely synchronized, the
percolation-based precursors successfully provide very
early warnings of the rapid switches between the two states of
the system. We clarify the mechanisms behind the
percolation transition by separating global trends given by the
mean-field behavior from the synchronization of individual
stochastic fluctuations. We then apply the same methodology to
real-world data of sea surface temperature anomalies during
different phases of the El Ni\~{n}o-Southern Oscillation. This
leads to a better understanding of the factors that make
percolation precursors effective as early warning indicators of
incipient El Ni\~{n}o and La Ni\~{n}a events.
\end{abstract}


\maketitle

\section{Introduction}
\label{sec:introduction}

The occurrence of sudden shifts between radically different
dynamical regimes is a striking phenomenon displayed by a
variety of complex systems, including climatic \cite{Alley2003,
Lenton2008,Thompson2011, Ashwin2012}, ecological
\cite{May1977,Scheffer2001}, financial \cite{Yan2010, Smug2018}
and physiological \cite{Maturana2020} ones. These so-called
critical transitions or regime shifts can have large impacts,
as they bring about a drastic change in the function and
structure of the systems undergoing them. Their forecast well
in advance is of utmost importance for risk management and the
mitigation of their impacts.

A large body of research has been devoted to the identification
of generic early warning signals for upcoming critical
transitions \cite{Scheffer2009, Scheffer2015, Lenton2011,
Kuehn2012, Romano2018, Kwasniok2018, Boers2021} and their
conceptualization is well developed for transitions induced by
bifurcations, accordingly named B-tipping in the nomenclature
of Ashwin~\emph{et al.} \cite{Ashwin2012}. In such situation, a
control parameter is changed gradually, until eventually a
critical threshold or tipping point is reached at which the
system suddenly reorganizes into a completely new dynamical
regime. Dynamical systems theory shows that this occurs as the
state of the system looses stability, which forces it to move
to a new attractor. In most cases \cite{Thompson2011}, as the
system approaches the point of loss of stability, the dominant
eigenvalue of its Jacobian matrix tends to zero, which
translates into decreasing relaxation rates towards equilibrium
when the system is perturbed \cite{Wissel1984, Wiesenfeld1986,
VanNes2007}. This phenomenon is known as critical slowing down
\cite{Strogatz2018}.

{A variety of early warning indicators for critical
transitions have been proposed based on statistical signatures
of critical slowing down, including temporal
\cite{Carpenter2006,Ives1995, Held2004,
Lenton2012,Guttal2008a,Kleinen2003,Seekell2011,Nazarimehr2017}
and spatial indicators \cite{Guttal2008b,
Dakos2011,Dakos2011,Guttal2008b, Dakos2011,Carpenter2010}. In
addition, some studies have integrated both spatial and
temporal indicators \cite{Tirabassi2022}.} Alongside these
developments, many studies have demonstrated the potential of
complex networks for the study of statistical structures in
systems consisting of interconnected units \cite{Nicolis2005,
Newman2010, NetsClimateBook2019, Fan2021, Ludescher2021}. In
particular, various works have used network-based frameworks
for the prediction of bifurcation-induced tipping
\cite{Mheen2013, Tirabassi2014, RodriguezMendez2016}. In
Rodriguez-Mendez~\emph{et al.} \cite{RodriguezMendez2016}
functional networks representing strong correlations between
different sites in spatially extended systems were studied and
different percolation quantifiers in such networks were shown
to provide useful anticipation of an incipient bifurcation.

{ More specifically, in systems consisting of
interconnected (e.g. spatially distributed) units coupled by diffusive or similar
types of homogenizing interactions, as a bifurcation is
approached, the variability between different units becomes
increasingly correlated as a consequence of the slowdown
mentioned above (there is an increasing amount of time for the
coupling to act). Such an increased coherence can
manifest in the dynamics at different timescales, from a
synchronization of fast fluctuations up to the emergence of
common long-term behaviors (e.g.\, stochastic trends). In a
correlation-based functional network this translates into an
increase of the link density. Hence, as conditions bring the
system closer to the bifurcation, the increasing connectivity
eventually results in the development of a large connected
component comprising a significant fraction of all nodes. This
is called percolation. Full connectivity of the network is
attained at the bifurcation but the percolation threshold is
actually reached \emph{before} the bifurcation. In this way,
percolation of the functional network is an early warning
signal of the approaching bifurcation, and precursor signals of
the percolation transition itself will provide even earlier
anticipation than standard anticipatory signals of the
bifurcation. The success of this approach was well illustrated,
for various B-tipping situations, in Rodriguez-Mendez \emph{et
al.}~\cite{RodriguezMendez2016}.}

Despite their generic nature and relevance across disciplines,
not all interesting or important sudden transitions in complex
systems are of the B-tipping type \cite{Ashwin2012,Feudel2018}.
For instance, the regime shift may be induced by noisy
fluctuations, not necessarily close to any bifurcation point,
making the system jump to an alternative dynamic attractor.
This is known as N-tipping. In other cases, the attractor
changes as the governing parameter is varied. If the rate of
change of the parameter is too fast, the system may leave the
basin of attraction, leading to so-called R-tipping. Further,
Kuehn \cite{Kuehn2011} considered tipping-like situations
induced by the bifurcation of the fast dynamics in two
timescale systems. Whether generic early warning signs of
sudden shifts can be found beyond the case of B-tipping remains
an open question and represents a significant challenge.

{ In this paper, we explore the ability of
percolation measures based on functional network
representations to anticipate sudden shifts in spatially
extended complex systems beyond the classical B-tipping
scenario, focusing specifically on systems presenting irregular
oscillations. We first consider fast-slow systems, choosing as
a leading example a system consisting of nonperiodic,
stochastic FitzHugh-Nagumo oscillators coupled by diffusion and
arranged in a one-dimensional ring lattice. All model
parameters remain fixed so that no bifurcation or B-tipping
occurs. {Nevertheless, jumps between distinct global
states occur at irregular times, for which we want to find
anticipatory signals.} To this end, we construct a
time-dependent functional network describing the evolution of
correlations of the system's fast variables at different
locations. {We show that percolation transitions occur in
the network, which anticipate the irregular sudden
changes between different stages of the system's oscillation}.
We then characterize the contributions of different processes
potentially causing the percolation transitions. These include
common trends among individual oscillators as well as the
synchronization of the superimposed (stochastic) fluctuations.
}

The obtained results {provide new insights into time-dependent structural changes emerging in functional network representations of multicomponent complex systems which}
lead to a better understanding and interpretation of the
percolation-based precursors in their application to real-world
oscillatory phenomena. {This is illustrated by
reconciling the previously reported anticipatory power of the percolation framework in
the climatic phenomena of El Ni\~{n}o and La Ni\~{n}a. Notably, } 
Rodriguez-Mendez \emph{et al.}~\cite{RodriguezMendez2016}
applied the percolation-based framework to time series of
sea-surface temperatures (SSTs) in the equatorial Pacific. They
showed the ability of the methodology to anticipate El Ni\~{n}o
and La Ni\~{n}a events, the two extreme stages of a single climate
oscillation: the El Ni\~{n}o-Southern Oscillation (ENSO). Other
works have also confirmed that different characteristics of the
percolation transition on climate networks provide early
warning indicators for the occurrence of El Ni\~{n}o or La
Ni\~{n}a \cite{Meng2017,Lu2016,Lu2018,Sonone2021}. In fact,
several topological as well as geometric properties of
functional climate networks have been reported to provide
useful diagnostics for the overall state of the spatially
extended coupled atmosphere-ocean system during different ENSO
phases \cite{Radebach2013, Kittel2021, Wiedermann2016}.
Nonetheless, the {mathematical} mechanisms leading to the percolation
transition in these correlation networks are far from being
fully understood. { In light of the understanding
gained from studying the processes causing the percolation
transitions in the FitzHugh-Nagumo system, we identify the
mechanism behind the performance of the percolation framework
for the anticipation of El Ni\~{n}o and La Ni\~{n}a reported in
Rodriguez-Mendez \emph{et al.}~\cite{RodriguezMendez2016}:  a
global tendency in the variation of SSTs across the considered
region. }

The remainder of this paper is organized as follows. In
Section~\ref{sec:network_construction} we describe the network
construction and the principles of percolation-based precursors
for critical transitions. Section~\ref{sec:FHN} first
introduces the system of coupled FitzHugh-Nagumo oscillators
and describes its numerical integration. The application of the
percolation-framework to this system and the results obtained
are then detailed. Finally, in
Section~\ref{sec:ENSO-percolation}, we show how the lessons
learned from the previous example shed light on the dynamical
processes inducing variations in the connectivity structure of
the (correlation-based) functional network describing El Ni\~no
and other phases of the El Ni\~{n}o--Southern Oscillation.

\section{Correlation networks and percolation}
\label{sec:network_construction}

\subsection{Functional networks}

Spatially extended dynamical systems can be approximated by
spatially localized units governed by their own local dynamics
while interacting with other units through processes such as
diffusion, convection, conduction, etc. Functional networks are
a discrete way of encoding statistical associations among these
individual local dynamics. They have been successfully employed
to get insights into the dynamics of complex systems in fields
as diverse as neurophysiology  \cite{Gerster2020}, urban
systems \cite{Neal2021}, seismology \cite{Abe2004} and
climatology \cite{Donges2009,
NetsClimateBook2019,Ludescher2021}.

In this paper, we analyze the emergence of {strong} correlations between
different locations of spatially extended dynamical systems by
means of a functional network constructed as follows. Let the
spatiotemporal field $u(x,t)$ describe (although perhaps only
in an incomplete manner) the evolution of the system's state,
where $x$ is space and $t$ is time. We discretize both space
and time to obtain the series $\{u_k(t_l)\}_{kl}$, where $k\in
\{1,\ldots,N\}$ labels the different spatial locations that
will be used as the $N$ nodes of the functional network, and
$l\in \{1,\ldots,R\}$ labels time instants. The links are built
through statistical analysis of inter-dependencies between
pairs of time series from different spatial locations. {Most commonly,} the
similarity between the variability at locations $a$ and $b$ is
computed via the zero-lag Pearson correlation
\begin{equation}
    \rho_{ab} = \frac{\sum_l p_a(t_l)p_b(t_l)}{\sqrt{(\sum_l p_a(t_l)^2)(\sum_l p_b(t_l)^2)}} \ ,
\label{eq:Pearson}
\end{equation}
where $p_k(t_l)=u_k(t_l) - \frac{1}{R}\sum_l u_k(t_l)$ is the deviation of the field from its temporal mean at each location. A link is set between the nodes $a$ and $b$ if the Pearson correlation is higher than a given threshold $\gamma$: $\rho_{ab} > \gamma$. {Only positive correlations are considered here, but it is straightforward to extend the formalism to the case $|\rho_{ab}| > \gamma$, as was done in Ekhtiari \emph{et al.} \cite{Ekhtiari2021} for instance.}

The (undirected) network constructed in this manner describes
the {strongest positive linear} correlations between parts of the system over the time
interval $[t_1,t_R]$. We call this the `network at time
$t_R$', stressing that it contains information only on the past
of time $t_R$, as adequate for a formalism that will be
intended for forecasting and anticipation. By repeating the
procedure over consecutive intervals $[t_2,t_{R+1}]$,
$[t_3,t_{R+2}]$, etc., we construct networks at times
$t_{R+1}$, $t_{R+2}$, etc., a sequence that conforms a
time-dependent functional network \cite{Radebach2013}, a
particular type of a temporal network \cite{Holme2012} whose
architecture reflects the evolution of the correlations of the
system {evaluated over time windows consisting of $R$ time steps each}. Hence, by analyzing how the network's topological
properties change over time, it is possible to detect a
reorganization of the correlation structure of the system at a
certain time, revealing changes in the spatiotemporal
dynamics.

{We note that the above network construction method is
just the simplest among the many available methodologies that use
network-theory frameworks to reveal dynamical properties of
spatially extended dynamical systems (see e.g.
\cite{NetsClimateBook2019}). Other measures of statistical
dependence, such as lagged
correlations\cite{Meng2017,Sonone2021}, or
information-theoretic measures \cite{Deza2015} can be used, which may lead to functional networks encoding different aspects of time series similarity \cite{Ferreira2021}. But here we
restrict ourselves to the simplest methodology, which can be applied
with essentially no knowledge on the underlying mechanisms or
on the spatial and temporal scales generating the
spatiotemporal signals. We will show that this framework is
powerful enough to find anticipatory signals of sharp
transitions, as well as to extract some information on their
origin. }

\subsection{Percolation}
Percolation theory {traditionally} describes the changes in network
connectivity as links are added {or removed} \cite{Newman2010,
Stauffer2018}. In many networks, provided that they are large
enough, there is a critical number of links at which the
disconnected clusters merge into a larger connected cluster
that contains a significant fraction of the nodes. Such
phenomenon is referred to as a percolation transition. Several
metrics can be used to characterize the cluster properties of a
network \cite{Newman2010}. In particular, we consider here the
\textit{{relative size of the largest connected component}},
$S_1$, i.e.\, the fraction of nodes that forms the largest
cluster, and the \textit{{probability that a {randomly
chosen} node belongs to a cluster of size $s$}}, $c_s$, $s \ne
1$. The later quantity can be {expressed} as
\begin{equation}
c_s= \frac{s n_s}{N},
\end{equation}
where $n_s$ is the number of clusters of size $s$ and $N$ is
the number of nodes in the network. Monitoring such
characteristics in the time sequence of evolving functional
networks allows to identify and potentially anticipate
percolation transitions in time-dependent networks.

{ Percolation measures from correlation networks have
been shown to provide very early warning signals of upcoming
B-tipping in spatially extended systems
\cite{RodriguezMendez2016}. Notably,
critical slowing down occurs close to various (although not all
\cite{Thompson2011}) types of bifurcations. {As already mentioned in the introduction,} in systems
consisting of interconnected spatial units with homogenizing
interactions (such as diffusive coupling), the
slowing down of the dynamics when approaching a bifurcation
gives ample time for the interactions to produce large
correlations, which in turn lead to a correlation network with
large link density. Eventually, the increasing connectivity
will result in a percolation transition in the functional
network. The percolation threshold is reached \emph{before}
(i.e. at lower connectivity than) the bifurcation (at which
full connectivity is attained). Thus, the occurrence of a
percolation transition as a system parameter is changed is
already an early warning signal announcing the proximity of a
bifurcation; and anticipatory signals of the percolation itself
will provide still earlier warning signals of the approaching
bifurcation or B-tipping. Rodriguez-Mendez \emph{et
al.}~\cite{RodriguezMendez2016} illustrate this approach for
various B-tipping situations. }

However, many important sudden shifts in nature cannot be
attributed to B-tipping processes. In the following, we
therefore investigate the potential of the percolation measures
for anticipating sudden shifts in spatially extended complex
systems beyond the case of B-tipping, focusing specifically on
systems presenting irregular oscillations. The aim will be to
anticipate associated abrupt changes in the system state that
occur during the oscillation cycles.

\section{Diffusively coupled noisy FitzHugh-Nagumo oscillators}
\label{sec:FHN}

\subsection{Description of the Model}

The FitzHugh-Nagumo (FN) system is a minimalistic and
prototypical model of an excitable system which can display an
oscillatory regime. It is a simplification of the
Hodgkin-Huxley model for the transmission of electrical pulses
along a nerve axon \cite{Hodgkin1952} which was first derived
by FitzHugh \cite{FitzHugh1955} and Nagumo \cite{Nagumo1962}.
The archetypal form of the model is given by

\begin{equation}
\begin{split}
&\dot{u} = f(u) - v + I, \\
&\frac{1}{\epsilon}\dot{v}=u-bv+c,
\end{split}
\label{eq:FHN}
\end{equation}
where $f(u) = \alpha u(u-a)(1-u)$ and $\alpha, a,  b, c, I$ are
constant parameters. The excitable variable $u$, corresponding
in the original model to the membrane's potential, is
characterized by a fast dynamics while the recovery variable
$v$ has a slow dynamics. The time scale separation between both
variables is given by $\epsilon \ll 1$. Depending on the choice
of parameters, system \eqref{eq:FHN} has one or three
stationary points, whose stability determines the solution's
behavior. For the set of parameters chosen in this study (which
includes $I=c=0$), the system exhibits an unstable focus located
at $(0,0)$ as well as a unique attractive limit cycle, leading
to periodic pulses. This type of
oscillations are of primary importance for the modeling of
neuro-biological systems \cite{FitzHugh1955, Nagumo1962,
Gerster2020}, in electrical circuits \cite{Bao2019}, in
seismology \cite{Cartwright1999}, and in climate science
\cite{Jin1997}.

In this study, we consider a system consisting of $N$
nonperiodic and stochastic FN-type oscillators coupled by
diffusion on a one-dimensional ring lattice.  The time
evolution of the state $(u_k,v_k)$ at node $k$ is described by

\begin{equation}
    \begin{split}
        &\dot{u_k} = f(u_k) - v_k + I + (\Delta u)_k + \sqrt{2D^{(u)}}\eta^{(u)}_k,  \\
        &\frac{1}{\epsilon}\dot{v_k} = u_k - b v_k + c + (\Delta v)_k + \sqrt{2D^{(v)}}\eta^{(v)}_k,
    \end{split}
    \label{eq:coupled_FHN}
\end{equation}
for $k=1,...,N$. We will refer to the label $k$ as ``space" in
the following. Node $k$ is coupled to its nearest neighbor on
each side via the diffusive terms  $(\Delta u)_k = u_{k+1} +
u_{k-1} -2u_k  $ and $(\Delta v)_k = v_{k+1} + v_{k-1} -2v_k $.
{To avoid unnecessarily complex dynamics we have chosen
the diffusion coefficients of the $u$ and $v$ variables to be
equal, and units of space are scaled so that this common
diffusion coefficient becomes unity. }The terms
$\eta_k^{(u)}(t)$ and $\eta_k^{(v)}(t)$ are two independent
Gaussian white noise sources, with $\langle \eta_r^{(a)}(t)
\eta_s^{(b)}(t')\rangle = \delta_{ab}\delta_{rs}\delta(t-t') $.
The addition of noise to the deterministic system is motivated
by the fact that real systems unavoidably include stochastic
fluctuations and by the necessity of well-defined spatial
correlation functions for the construction of the correlation
network described in Section \ref{sec:network_construction}.
Despite the presence of these additive noise terms, if their
intensity remains small, the resulting FN dynamics is a rather
regular periodic oscillation. In order to bring the model
closer to the type of irregular oscillations that we want to
address {--for which prediction is a non-trivial task--} we let the parameter $\epsilon$,
{which has the most direct influence on} the
timescale separation between $u$ and $v$, vary stepwise in
time. Precisely, $\epsilon$ takes a constant value during the
time needed for the system to complete a full oscillation
cycle. In this case, a full oscillation cycle is defined as a
full cycle of the mean dynamics over all $N$ oscillators.
 Once a cycle has been completed, a new value of $\epsilon$ is chosen
(the same for all $k$) with uniform probability over the
interval $(0.001, 0.033)$. {The range is chosen such as
to produce oscillations with period lengths suited to the
purpose of our study.} This choice of $\epsilon=\epsilon(t)$
introduces a strong irregularity in the period of the
oscillations making the system more {similar to}
natural oscillatory phenomena, and making the anticipation of
the different abrupt changes substantially more challenging.

\subsection{Numerical integration}

In the following, we choose  $\alpha = 1, a=-1, b = -0.83, c =
0, I = 0$  and noise intensities $D^{(u)}= D^{(v)}=0.01$ . We
consider $N=500$ oscillators.

To compute numerically the solution to System
\eqref{eq:coupled_FHN}, we start from a uniform initial state
$(u_k(0)=-0.16,v_k(0)=0.01)$ $\forall\ k\in\{1,...,N\}$. The
solution is evolved with an integration time step $dt=0.05$. At
each time step the deterministic part of the equation is
integrated first via a fourth order Runge-Kutta scheme
\cite[sect. 7.4]{Toral2014}. Then, $N$
independent random Gaussian distributed numbers of zero
mean and variance $2D^{(u)}dt$ and $2\epsilon^2D^{(v)}dt$ are added to
$u_k$ and $v_k$, $k\in\{1,...,N\}$, respectively.

\begin{figure}
\centering
\includegraphics[trim={2cm 0 2cm 0}, width = 8.6cm]{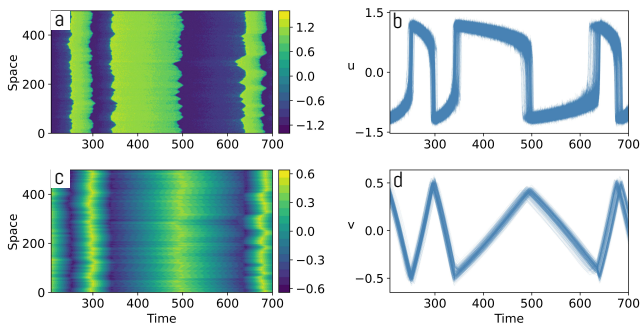}
\caption{\emph{Left}: evolution of the fields $u$ (a) and $v$
(c) as a function of space and time. \emph{Right}: all ($N=500$) individual time series superimposed for $u$ (b) and $v$ (d).}
\label{fig:FHN-system}
\end{figure}

Figure \ref{fig:FHN-system} shows the evolution of the solution
$(u,v)$. Figures \ref{fig:FHN-system}(a) and \ref{fig:FHN-system}(c) show the evolution of the
fields $u$ and $v $ as a function of space and time.
As an alternative view, Figs. \ref{fig:FHN-system}(b) and
\ref{fig:FHN-system}(d) show the individual time series of $u$ and $v$
superimposed for all ($N=500$) nodes. The evolution of the
variables $u$ and $v$ is quite coherent in space, i.e. all
nodes $k$ are nearly perfectly synchronized, a consequence of
the diffusive coupling and of the choice of parameters. We
observe that the state variable $u$ presents asymmetric and
nonperiodic oscillations with sudden switches between
positive/negative values. The state variable $v$ shows linear
increase and
decrease in time.

In the next section, we investigate the potential of the
percolation precursors to anticipate the irregular abrupt
global changes in the state of the fast variable $u$, without
using any information on the variable $v$. To do so, we
construct a sequence of networks from the discrete field
$u_k(t_l)$ following the methodology described in the previous
section, using consecutive time windows of 140 steps, that is
of 7 time units {($140\cdot dt$)}, and a {correlation} threshold $\gamma = 0.6$.
Information from the field $v_k(t_l)$ is ignored in the network
construction. As mentioned previously, the network (and network
quantities) at time $t$ describes the information contained in the time window prior to $t$.

Note that the predictability of the behavior of $u$ is made
significantly more challenging by (1) the time
dependence of $\epsilon$ which leads to the irregularity of the
oscillations, as well as (2) the absence of information
from the field $v$ in the network construction.

\subsection{Percolation-based precursors in the FN system}
\label{subsect:FNprecursors}

\begin{figure}
\centering
\includegraphics[trim={0.2cm 1cm 0.2cm 0}, width = 8.6cm]{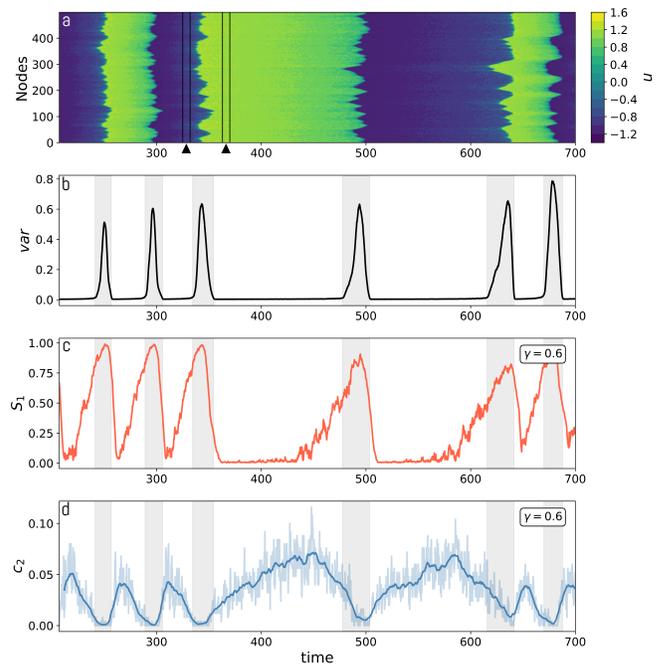}
\caption{\emph{(a)} Spatiotemporal evolution of the variable $u$,
\emph{(b)} spatial variance of $u$,
\emph{(c)} relative size $S_1$ of the largest connected component in the network constructed
from the variables $u_k$ with correlation threshold $\gamma = 0.6$,
\emph{(d)} probability $c_2$ that a randomly chosen node belongs to a component of size $2$ in the
network. The light blue line gives the raw values of $c_2$ whereas the darker curve smooths the data using a
$40$-point window average (i.e. $2$ time units), centered at the
time of interest. The gray shading indicates the time interval during which the $u_k$s switch sign. The two triangles indicate the locations of two time intervals of 140 time steps each whose  correlation properties are displayed in Fig. \ref{fig:correlation-structures}.}
\label{fig:non-detrended-percolation-measures-FHN}
\end{figure}

The increase in spatial variance is commonly used as a
precursor of abrupt changes in dynamical systems
\cite{Guttal2008b, Dakos2011}. Therefore, we first compute the
variance of the state variable $u$ as $\bar{\sigma}_u(t_l) =
\sqrt{\frac{1}{N}\sum_{k=1}^N (u_k(t_l)-\bar{u}(t_l))^2}$,
where $\bar{u}(t_l) = \frac{1}{N}\sum_{k=1}^Nu_k(t_l)$ is a
spatial average (over all the nodes). Figure
\ref{fig:non-detrended-percolation-measures-FHN}(a) shows again
the evolution of the $u_k$ time series at all nodes for reference.
The gray shading in Figs. \ref{fig:non-detrended-percolation-measures-FHN}(b), \ref{fig:non-detrended-percolation-measures-FHN}(c) and
\ref{fig:non-detrended-percolation-measures-FHN}(d) indicates the time intervals during which the values
of $u$ are changing sign, the bounds of this interval being
defined by the first and the last oscillator to cross zero. In
Fig. \ref{fig:non-detrended-percolation-measures-FHN}(b) we can see that the variance peaks within this
interval. However, there is no detectable rise of spatial
variance until the system enters the "jump" time interval, i.e. until
the first oscillators switches sign. Hence, for this system, the
peak in variance indicates the occurrence of the abrupt change, but it is not useful as an anticipatory or
early warning tool.

Figures \ref{fig:non-detrended-percolation-measures-FHN}(c) and \ref{fig:non-detrended-percolation-measures-FHN}(d) show
the evolution of the network measures $S_1$, the relative size
of the largest connected component, and $c_2$, the probability
that a randomly chosen node belongs to a component of size $2$
\cite{RodriguezMendez2016}. Note that in Fig. \ref{fig:non-detrended-percolation-measures-FHN}(d) the
light blue curve shows the actual values of $c_2$ while the
dark blue curve gives a smoothing of the data which makes the
general increase and decrease in the values of $c_2$ more
apparent. The smoothing is obtained using a $40$-point window
average (the equivalent to $2$ time units), centered at the
point of interest.

We observe that both $S_1$ and $c_2$ anticipate the abrupt
increases and decreases in the state variable associated with
the system's oscillations. Precisely, $S_1$ starts increasing
(indicating the start of the percolation phase) long before the
sharp changes in the values of $u_k$. { It reaches
its maximum during the transition of the state variable and
then decreases very rapidly, reaching $0$ just after the
transition. The behavior of $S_1$ reflects that the
correlation between the dynamics at individual nodes increases
as a jump in the $u$ values is approached and} accordingly the
link density in the network also increases so that eventually a
cluster of significant size forms. The numbers of nodes in
clusters of small sizes (2, 3, 4, etc.) first increase, before
decreasing again as most of the nodes begin to attach to the
percolating cluster of relative size $S_1$. This leads to the
presence of a peak in the values of $c_2$ just as $S_1$ starts
increasing [Fig.
\ref{fig:non-detrended-percolation-measures-FHN}(d)],
providing a very early warning signal of the abrupt change in
$u$. Note that the asymmetry of $S_1$ around the jump and the
lack of peaks in $c_2$ after the shift are due to the
irreversibility of the system's abrupt jumps: the correlation
state in the network changes completely after $u_k$ has
switched sign for all oscillators.

Despite the irregularity of oscillations and the
fact that information from the variable $v$ was ignored in the
network construction, the network measures are well able to
anticipate the abrupt changes in the values of $u$. However, we saw that the spatial variance did not provide
significant anticipation. This shows the advantage of using the
percolation measures as precursors of abrupt changes in the
present system.

\subsection{Processes leading to percolation}

The occurrence of percolation in the system's functional
network, manifested in the increase in $S_1$ before the abrupt
global jump in the oscillator's state occurs, reflects an
increase in correlations which can be triggered by two
different processes {(or by a combination of the two)}:
\begin{enumerate}
\renewcommand{\theenumi}{(\roman{enumi})}
\item From the definition of the spatial Pearson
    coefficient in Eq. (\ref{eq:Pearson}) it is clear that
    a large and coherent change {common to} all
    spatial units gives a much larger value of $\rho_{ab}$
    than small fluctuations around some stationary base
    state. Thus, in systems for which an abrupt change in
    the global state of the system is preceded by an
    upward/downward trend of all spatial units, percolation
    in the correlation network will occur giving an early
    warning of the {incipient} sudden jump.  This is a rather
    general mechanism, that should provide anticipatory
    signals in a variety of real systems in which sharp
    global changes interrupt relatively quiescent states.

\item In the particular case of the FN system, an
    additional mechanism could be at work. As already
    mentioned, $u$ and $v$ evolve on two different time
    scales. From the structure of the local equation
    (\ref{eq:FHN}), the 'slow' variable $v$ can be seen as
    a bifurcation parameter that drives the dynamics of the
    'fast' variable $u$ \cite{Meron1992}: $u$ drifts slowly
    until $v$ crosses a critical threshold, which produces
    an effect similar to a saddle-node bifurcation of the
    $u$ dynamics. As a consequence, a rapid shift occurs to
    an alternative $u$-state. In fact, one can clearly
    anticipate the jumps in $u$, even if irregular, by
    looking at the trend in $v$ and estimating when it will
    reach a critical value (see Fig. \ref{fig:FHN-system}).
    This is why we have ignored the information about $v$
    in the network construction, to be closer to real-world
    situations in which relevant variables may not be
    directly observable. {The situation here} is very
    similar to the B-tipping described earlier. The
    critical slowing down associated to the effective
    bifurcation in $u$ can thus explain the increase in
    correlations and the percolation anticipating every
    sudden jump. This second mechanism is solid and theoretically well understood,
    but is not as general as the first one, as it will only
    be present in fast-slow dynamical systems {(which are indeed commonly used to model critical
    transitions \cite{Kuehn2011})}.
\end{enumerate}

{We aim to elucidate which of the two mechanisms above
(or which combination of them) is responsible for the success
of the percolation framework in anticipating the FN system
transitions. To this end, we propose to repeat the percolation
analysis with modified spatiotemporal data in which all the
synchronous upward/downward trends have been removed, and thus
any remaining anticipatory power should arise from mechanism
(ii) alone. More specifically,} we remove the upward/downward
trends from the time series at each location $k$ in our coupled
FN model by using a Gaussian kernel. That is, for each time
step $t_l$ of the time series at $k$, $u_k(t_l)$, we remove the
weighted average:
\begin{equation}
    \frac{1}{\sum_{l'} \omega_{l'}^{(l)}} \sum_{l'} \omega_{l'}^{(l)} u_k(t_{l'}) \ ,
    \label{eq:Gaussian-smooting}
\end{equation}
where $\omega_{l'}^{(l)} = \exp[-(t_l-t_{l'})^2/(2b^2)]$ and the
bandwidth is $b=40$ (corresponding to $2$ time units).
{ The time window is centered {around $t_l$}
to remove trends in the best possible way. {This implies
taking into account some future values to obtain the detrended
time series at $t_l$. This is not a problem since in this
subsection we are only interested in elucidating the mechanism
for the success of the percolation early warnings, and for this
we are postprocessing the whole time series. The finding of the
anticipatory signals themselves was properly done in Section
\ref{subsect:FNprecursors} by using only past values of the
analyzed time series, as appropriate for real-time monitoring
applications}. }

\begin{figure}
\centering
\includegraphics[trim={0.2cm 1cm 0.2cm 0}, width = 8.6cm]{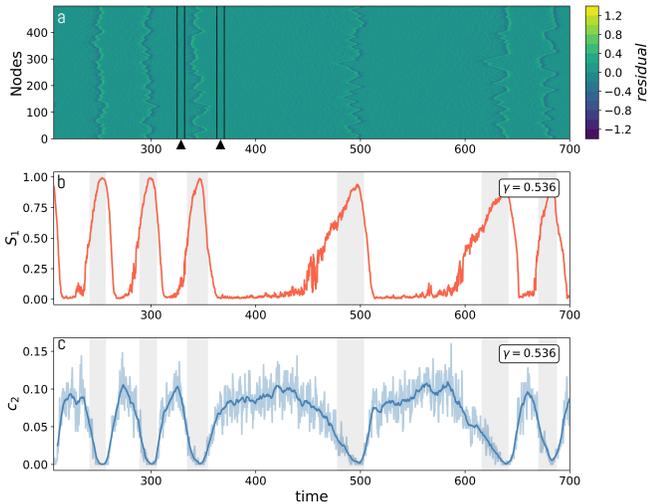}
\caption{\emph{(a)} Residual fluctuations obtained after detrending
the variables $u_k$ , \emph{(b)} relative size $S_1$ of the largest connected component
in the network constructed from the detrended time series for $u_k$ with correlation threshold
$\gamma = 0.536$, \emph{(c)} probability $c_2$ that a randomly chosen node belongs to a component
of size $2$. The light blue line gives the raw values of $c_2$ whereas the darker curve smooths the data using a
$40$-point window average (i.e. $2$ time units), centered at the
time of interest. The gray shading indicates the time interval during which the $u_k$s switch sign. The two triangles indicate the locations of two time intervals of 140 time steps each whose  correlation properties are displayed in Fig. \ref{fig:correlation-structures}.}
\label{fig:detrended-percolation-measures-FHN}
\end{figure}

The residual variability is shown in Fig.
\ref{fig:detrended-percolation-measures-FHN}(a). The
time-series of $S_1$ and $c_2$ for the network constructed on
the detrended fluctuations are shown in Figs.
\ref{fig:detrended-percolation-measures-FHN}(b) and \ref{fig:detrended-percolation-measures-FHN}(c). The
gray shadings bound the jump regions where transient effects
remain due to large variations in the data which impede
complete detrending. Because of the detrending, mechanism (i)
should be absent, and only mechanism (ii) should be at work, at
least outside the gray shaded regions. We observe that the
anticipation power is similar to the one obtained using the
original (i.e. non-detrended) data. This result suggests that
critical slowing down indeed takes place in this system before
every sudden shift, leading to increased spatial correlations.
Note, however, that the chosen correlation threshold is now of
$\gamma = 0.536$ versus $\gamma = 0.6$ in the above example. In
fact, the strength of the correlations between spatial points
should be generally lower when the network is constructed from
detrended data, that is when mechanism (i) is absent.

\begin{figure}
\centering
\includegraphics[width = 8.9cm]{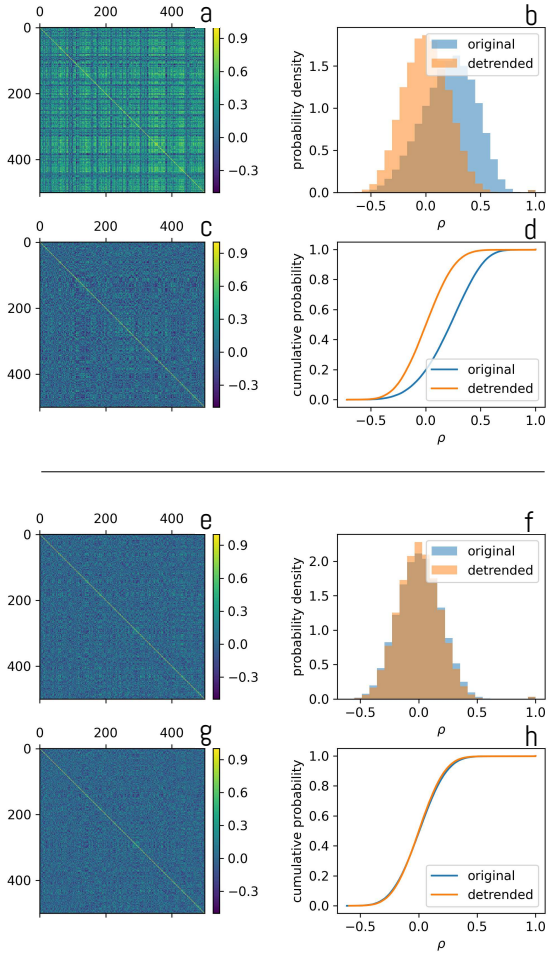}
\caption{\emph{a-d}: Correlations computed from the original and detrended data
over the time interval $[325,332]$ indicated by the first black triangle in
Figs. \ref{fig:non-detrended-percolation-measures-FHN} and \ref{fig:detrended-percolation-measures-FHN}.
Panels \emph{a} and \emph{c}: the correlation matrices computed from the original and
detrended data respectively. Panels \emph{b} and \emph{d}: comparison of the correlation
distributions (probability density and cumulative probability) for the two cases. Panels
\emph{e-h}: correlations over the time interval $[363,370]$ indicated by the second black
triangle in Figs. \ref{fig:non-detrended-percolation-measures-FHN} and
\ref{fig:detrended-percolation-measures-FHN}. Panels \emph{e} and \emph{g}: correlation
matrices computed from the original and detrended data.  Panels \emph{f} and \emph{h}:
comparison of the correlation distribution and the cumulative distributions.}
\label{fig:correlation-structures}
\end{figure}

Figure \ref{fig:correlation-structures} further illustrates the
effect of mechanism (i) on the correlation values by inspecting
their distributions over two different time intervals during
the evolution of the system, characterized, respectively, by
high [Figs. \ref{fig:correlation-structures}(a)-\ref{fig:correlation-structures}(d)] and low [Figs \ref{fig:correlation-structures}(e)-\ref{fig:correlation-structures}(h)] common
trends among the individual dynamics. First, Figs.
\ref{fig:correlation-structures}(a)-\ref{fig:correlation-structures}(d) characterize the
system's correlations over the time interval $[325,332]$ (see
first marker in Figs.
\ref{fig:non-detrended-percolation-measures-FHN} and
\ref{fig:detrended-percolation-measures-FHN}), that is just
before a jump in the values of $u$. The correlation matrices
for the original and detrended data are plotted in Figs. \ref{fig:correlation-structures}(a) and \ref{fig:correlation-structures}(c), respectively. The higher correlation
values and more complex structure of the non-detrended case [Fig. \ref{fig:correlation-structures}(a)]
are evident. In order to get a more objective
comparison of the correlation values, we show the probability
density of values in Fig. \ref{fig:correlation-structures}(b) for both cases as well as
their cumulative distributions in Fig. \ref{fig:correlation-structures}(d). These confirm
the previous observation: the correlations computed from the
original data are overall much higher than the ones computed
from the detrended data. Thus, in addition to the
fluctuation synchronization observed in the detrended data
[mechanism (ii)], there is also a large global trend common to
the dynamics of all spatial units in the non-detrended case [mechanism (i)]
when the system is close to a transition.

This interpretation is further confirmed by Figs.
\ref{fig:correlation-structures}(e)-\ref{fig:correlation-structures}(h), which display the
same quantities as Figs.
\ref{fig:correlation-structures}(a)-\ref{fig:correlation-structures}(d) but over the the
time interval $[363,370]$. This is just after the jump and
relatively far from the next one (see second marker in Figs.
\ref{fig:non-detrended-percolation-measures-FHN} and
\ref{fig:detrended-percolation-measures-FHN}). In this
interval, the upward/downward trend of all spatial units is not
as marked and little differences are observed in the structures
and distributions of correlations between the original and
detrended case.

Note that the difference in correlation strengths between the
original data and detrended data illustrated in Figs.
\ref{fig:correlation-structures}(a)-\ref{fig:correlation-structures}(d) clearly affects the
link density in the network. In fact, for any choice of
{the correlation} threshold $\gamma$, a {larger} fraction of
correlation values $\rho_{ab}$ will satisfy the criterium
$\rho_{ab}>\gamma$ when these correlations are computed from
the original data than from the detrended data. This leads to
higher network connectivity in the former case and the choice
of different thresholds $\gamma$ in Figs.
\ref{fig:non-detrended-percolation-measures-FHN} and
\ref{fig:detrended-percolation-measures-FHN}.\hfill \break

The results presented in this section demonstrate that the rise
of the percolation measures in the network computed from the
original data (reflecting a rise of correlation in the
{FN} system) is {supported} by both processes (i) and
(ii), {although the presence of mechanism
(ii) is powerful enough on its own to provide clear anticipatory signals
(Fig. (\ref{fig:detrended-percolation-measures-FHN})).}

{Distinguishing the contribution from each process allows
us to better understand the behavior of the percolation
precursors and the situations in which they are useful beyond
the case of the FN system. Specifically, the percolation
measures have the potential to detect a global trend in the
dynamics of any extended system very early on.  Thus, given the
knowledge that an increasing/decreasing global trend precedes
an abrupt change in the state of the system, the sensitivity of
the percolation measures can be leveraged to design very early
warnings of upcoming shifts. Moreover, in systems in
which some slowing down of the dynamics occurs before the
jumps, the anticipatory signals become still more powerful. A
postprocessing of the spatiotemporal series helps to reveal
the main mechanism of the percolation transition, thus
informing about system dynamics.  This later point is
illustrated further in the next section using a {real-world climate example}.}

\section{Percolation in sea surface temperature networks during
El Ni\~{n}o and La Ni\~{n}a events}
\label{sec:ENSO-percolation}

El Ni\~{n}o and La Ni\~{n}a are the hot and cold phases of
ENSO, the dominant interannual oscillation in the tropical
Pacific, which affects the whole Earth climatic system
\cite{Dijkstra2013}. The average ENSO period is about 4 years,
but with strong variability which makes its prediction
challenging \cite{Nooteboom2018, Dijkstra2019}.

Several studies, using different data sets, variables, network
construction methods and percolation indicators
\cite{RodriguezMendez2016,Meng2017,Lu2016,Lu2018,Sonone2021}
demonstrate that a percolation transition often occurs
in climatic variables of the tropical Pacific before El
Ni\~{n}o and La Ni\~{n}a events. However, there is no
obvious parameter change during the ENSO cycle that could
explain the performance of these indicators as arising from the
crossing of a bifurcation or B-tipping. In fact, while a Hopf bifurcation is
present in {some} models of ENSO \cite{Dijkstra2013}, such that its
dynamics can be conceptualized either as subcritical
oscillations excited by noise, or supercritical noisy
oscillations, the sharp changes that define the El Ni\~{n}o
and La Ni\~{n}a events themselves cannot readily be identified as the
crossing of any bifurcation.

In the previous section, we showed that, when the system of
coupled FN oscillators approaches an abrupt change, the spatial
correlations increase and that two combined factors could
{lead to} this increase, namely the effect of a common trend in
the monitored variables, or the impact of slowing down of the
dynamics when approaching the event. In light of these
observations, we now investigate the following question: ``Is the
'ENSO-percolation' related to a large-scale trend in the tropical
climatic variables or due to critical slowing down on
particular phases of the dynamics?".

\begin{figure}
    \centering
    \includegraphics[trim={0cm 8cm 0cm 9cm}, width = 8.6cm]{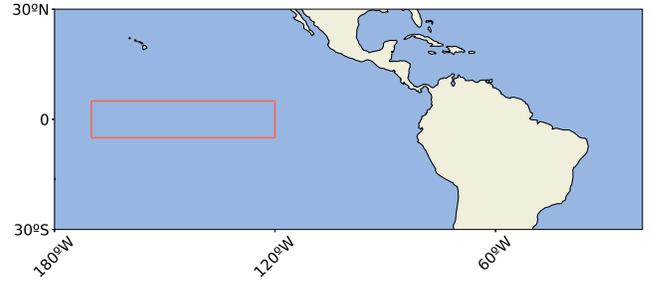} %
    \caption{The red box indicates the region (5S-5N/120W-170W) over which the mean sea surface temperature is monitored in the NINO3.4. index. The nodes of the functional network are chosen on a regular $0.375^\circ \times 0.375^\circ$ grid in this region (see text).}
    \label{fig:domain}
\end{figure}

To answer this question, we build on the network studies of
\cite{RodriguezMendez2016} and compare the evolution of the
percolation measures for a network computed from sea
surface temperatures (SSTs) over the tropical Pacific and from
the detrended SSTs over the same region. Precisely, the SST
data is obtained from the European Center for Medium-Range
Weather Forecast (ECMWF), ERA-Interim reanalysis product
\cite{Dee2011}, a global data set, covering a period of 40
years: from January 1979 to August 2019. We choose $3618$
network nodes lying on a $0.375^\circ \times 0.375^\circ$ grid
in the region 5S-5N/120W-170W - that is the region of the
Pacific used in the computation of the commonly employed
NINO3.4 index. The domain is illustrated in Fig.
\ref{fig:domain}. At each node, daily SST anomalies are
computed by subtracting the mean seasonal cycle. The links of
the network are set based on the correlation of the anomaly
time series over a sliding window of $200$ days resulting in a
time-dependent network.

\begin{figure*}
\centering
\includegraphics[trim={2cm 0 2cm 0}, scale = 0.45]{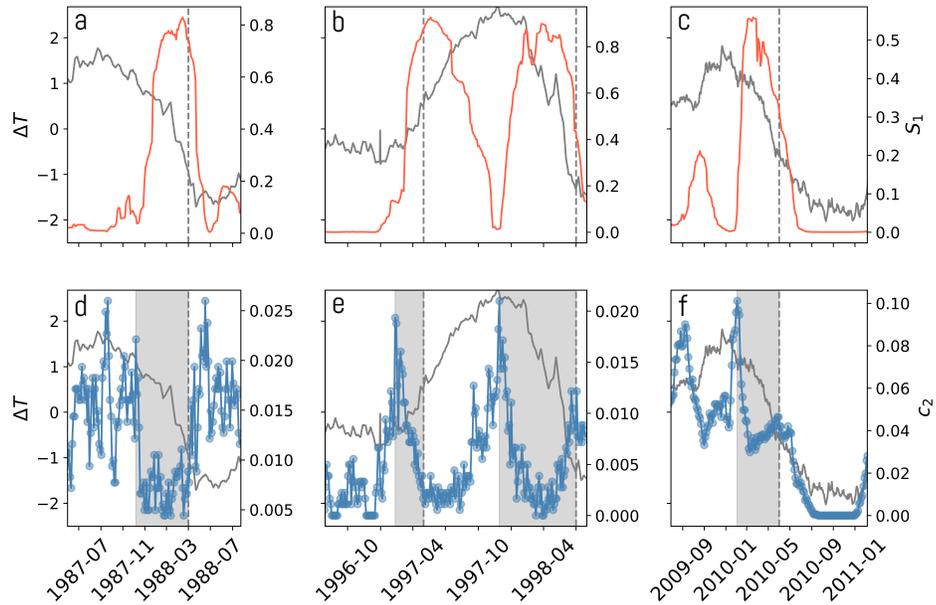}
\caption{\emph{a,b,c} Average SST anomaly over the region
5S-5N/120W-170W, for three different periods (gray). The dashed vertical lines
mark the beginning of El Ni\~{n}o/La Ni\~{n}a events.
Relative size of the largest connected component
$S_1$ (red) in the correlation network of SST built with {correlation}
threshold $\gamma=0.999$ for the first two periods and $\gamma
= 0.998$ for the last period. \emph{d,e,f} Probability $c_2$
that a randomly chosen node belongs to a component of size 2 in
the correlation network of SST (blue). Reproducing of results from
Rodriguez-Mendez \emph{et al.} \cite{RodriguezMendez2016}
} \label{fig:ENSO-percolation-original}
\end{figure*}

\begin{figure*}
\centering
\includegraphics[trim={2cm 0 2cm 0}, scale =0.45]{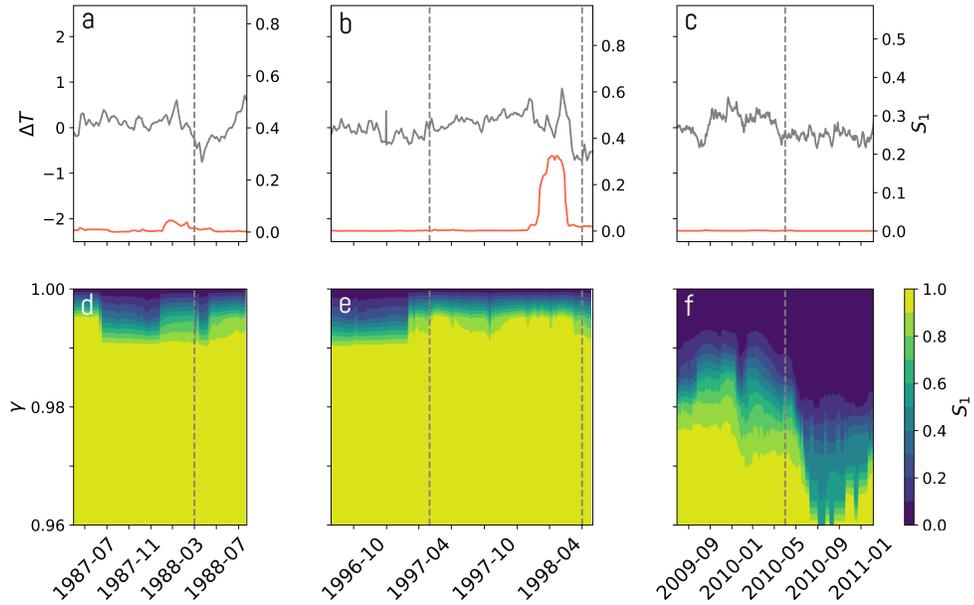}
\caption{\emph{a,b,c} Average \emph{detrended} SST anomaly over the region 5S-5N/120W-170W (gray).
The dashed vertical lines mark the beginning of El Ni\~{n}o/La Ni\~{n}a
events. Relative size of the largest connected
component $S_1$ (red) in the correlation network of \emph{detrended} SSTs
built with {correlation} threshold $\gamma = 0.999$ for the
first two periods and $\gamma = 0.998$ for the last period.
\emph{d,e,f} Values of $S_1$ for a range of
{correlation} thresholds $\gamma \in [0.96, 1.00]$.}
\label{fig:ENSO-percolation-detrended}
\end{figure*}

We focus on three periods: July 1987- July 1988, October 1996 -
April 1998 and September 2009 - January 2011. The mean SST
anomalies over the considered region are indicated by the gray
curves in Fig. \ref{fig:ENSO-percolation-original}. The
abrupt drops and increases in temperature correspond to La
Ni\~{n}a and El Ni\~{n}o events, respectively.  These events
are commonly defined by five consecutive 3-month running mean values
of sea surface temperature (SST) anomalies in the Ni\~{n}o 3.4
region below (above) the threshold of $-0.5^\circ$C
($+0.5^\circ$C). The starting date, i.e. the first day of the
first month, of each event is indicated by a vertical dashed
line in the figure.

The relative size of the largest connected
component of the network computed from the SST anomalies is
shown in Figs. \ref{fig:ENSO-percolation-original}(a)-\ref{fig:ENSO-percolation-original}(c) (red curves) for correlation threshold $\gamma =
0.999$ for the 1988, 1997 and 1998 events and $\gamma = 0.998$
for the 2010 event. The high correlation thresholds are necessary here to prevent the network from being fully connected and hence to observe changes in the network measures $S_1$ and $c_2$. We clearly see the increase in $S_1$ right
before every event as reported in \cite{RodriguezMendez2016}.
In addition, Figs. \ref{fig:ENSO-percolation-original}(d)-\ref{fig:ENSO-percolation-original}(f) (blue curves) show the probability for a randomly chosen node
to belong to a component of size $2$, $c_2$. This quantity
provides additional anticipation indicated by the gray shading
in the figure. Precisely, the anticipation provided by $c_2$ is
of 146 days for the event in 1988, 79 days for the event in 1997,
214 days for the event 1998 and 119 days for the event in 2010.

{Next, we want to identify which is the mechanism --
global trend or fluctuation synchronization by critical slowing
down-- which leads to the observed percolation transition. To
this end, as in the case of the FN oscillators, we postprocess
the spatiotemporal series to detrend them} at each grid point
using a Gaussian kernel with a bandwidth of 60 days. The
detrending is illustrated in Figs.
\ref{fig:ENSO-percolation-detrended}(a)-\ref{fig:ENSO-percolation-detrended}(c) by showing, in
gray, the spatial mean of the fluctuations for each considered
period. Figures \ref{fig:ENSO-percolation-detrended}(a)-\ref{fig:ENSO-percolation-detrended}(c) also show the time series of $S_1$
(in red) for the network computed from the detrended data for
the same correlation thresholds as above. We observe the lack
of peak in $S_1$ [except perhaps in Fig. \ref{fig:ENSO-percolation-detrended}(b)]. In order
to make sure that the absence of signal in $S_1$ is not due to
the choice of $\gamma$, we show the values of $S_1$ for a range
of thresholds in Figs. \ref{fig:ENSO-percolation-detrended}(d)-\ref{fig:ENSO-percolation-detrended}(f). In all cases the lack of
clear structures in the measure confirms the lack of variation
in the correlation of the fluctuations preceding ENSO episodes.
We note however that a drop in correlations occurs after the La
Ni\~{n}a event in the 2009-2011 period, which does not have an
anticipatory value. A more detailed analysis would be required
to identify its causes.

We can thus conclude that the increase in the size of the
largest connected component preceding El Ni\~{n}o and La
Ni\~{n}a events is due to a large-scale trend in the SSTs before
every event, and not to a slowing down of the dynamics at some
locations in space. This is relevant to discard possible
mechanisms, as some models of ENSO {obey} a slow-fast structure
very similar to the FN model \cite{Jin1997}. The percolation
precursors prove to be powerful tools to monitor the ENSO
irregular oscillation, being sensitive enough to detect the
{emergent} trend very early on and hence allowing for very early
warning signals of approaching El Ni\~{n}o and La Ni\~{n}a
events.

\section{Conclusion}
In summary, {we have analyzed} time-dependent functional
networks encoding the evolution of correlations in spatially
extended systems. The development of structures in such
networks allows to detect subtle changes in the systems
dynamics which might {easily be overlooked} by other analysis
techniques. Precisely, we saw that the percolation precursors
anticipate the abrupt changes between different stages of the
oscillation in a system of coupled stochastic FitzHugh-Nagumo
oscillators. The percolation reflects an increase in the
correlations of the system which is due to the combination of
two processes: (i) a global trend or coherent tendency among
all spatial units preceding the abrupt change and (ii) an
increase in the correlation of the noisy fluctuations ``on top
of" that trend (critical slowing down).

These considerations lead to a better understanding and
interpretation of the percolation precursors in their
application to the El Ni\~{n}o-Southern Oscillation. Our
results demonstrate that the increase in correlation of sea
surface temperatures over the NINO3.4 region preceding the studied El
Ni\~{n}o and La Ni\~{n}a events is triggered by a large-scale trend
in the SSTs, {that is a growth or decay of the SSTs
at all{, or at least sufficiently many,} spatial points,} before every event and not by a slow
down of the temperatures dynamics. This provides insights on
the mechanism causing the performance of the percolation
precursors in this particular case. { We focused on
a local scale, considering the El Ni\~{n}o basin only, and used
instantaneous Pearson correlations in the network construction. Hence, our
conclusions cannot be directly transferred to other percolation
studies of El Ni\~{n}o which were performed on larger scales
and used lagged-correlations such as in
\cite{Meng2017,Sonone2021}. In those cases, the percolation is
marked by the appearance of large-scale clusters and
teleconnections, for which the causing factor remains to be
found.}

 Tracking the structural changes in network descriptions of complex systems have the potential to
reveal various dynamical aspects of these systems. Beyond the
case of El Ni\~{n}o, these approaches have been utilized by
Gupta \emph{et al.} to track and identify short lived tropical
cyclones \cite{Gupta2021}. Sun \emph{et al.} \cite{Sun2021}
implemented a percolation model to investigate the structure
 and complexity of the atmosphere. Fan \emph{et al.} \cite{Fan2018} have used a
complex network framework to investigate changes in the
atmospheric circulation associated with global warming and
evaluate the impacts of future climate change. These examples
suggest the potential of complex networks and their percolation
properties to study present and future climate. Along this
line, our work brings further insights into the situations in
which the percolation-based precursors may be useful.

To facilitate future usage of the percolation framework
described in this work, we provide in \cite{github} a Python
notebook performing the integration of the FitzHugh-Nagumo
system with parameters specified in Section \ref{sec:FHN} and
computing the correlation network as well as its percolation
measures (Section \ref{sec:network_construction}).

\section*{Acknowledgement}
This project has received funding from the European Union’s Horizon 2020 research
and innovation programme under the Marie Skolodowska-Curie Grant Agreement No 813844,
from the Agencia Estatal de Investigación (MCIN/AEI/10.13039/501100011033)
under the María de Maeztu project CEX2021-001164-M, and from the Agencia
Estatal de Investigación (MCIN/AEI/10.13039/501100011033) and FEDER “Una
manera de hacer Europa” under Project LAMARCA No. PID2021-123352OB-C32.
R.V.D. acknowledges financial support by the German Federal Ministry for Education and Research (BMBF)
via the JPI Climate/JPI Oceans Project ROADMAP (Grant No. 01LP2002B).

\FloatBarrier

\providecommand{\noopsort}[1]{}\providecommand{\singleletter}[1]{#1}%

\end{document}